\documentclass[sn-standardnature,iicol]{sn-jnl}

\usepackage[compact]{titlesec}
\usepackage{caption}
\usepackage{makecell}
\titleformat{\section}{\normalfont\normalsize\bfseries}{0pt}{}{}
\titleformat{\subsection}{\normalfont\small\bfseries}{0pt}{}{}
\jyear{2021}%

\begin{document}

\title[Article Title]{Precise control of crystallography and magnetism in focused-ion-beam transformed iron-nickel thin films}


\author*[1]{Jakub Holobr\'{a}dek} \email{holobradek@vutbr.cz}

\author[1]{Libor Voj\'{a}\v{c}ek}

\author[1]{Ond\v{r}ej Wojewoda}

\author[3]{Michael Schmid}

\author*[1,2]{Michal Urb\'{a}nek}  \email{michal.urbanek@ceitec.vutbr.cz}

\affil[1]{\orgdiv{CEITEC BUT}, \orgname{Brno University of Technology}, \orgaddress{ \city{Brno}, \country{Czech Republic}}}

\affil[2]{\orgdiv{Institute of Physical Engineering}, \orgname{Brno University of Technology}, \orgaddress{\city{Brno}, \country{Czech Republic}}}

\affil[3]{\orgdiv{Institute of Applied Physics}, \orgname{TU Wien},  \city{Vienna}, \country{Austria}}


\abstract{Focused ion beam irradiation of metastable Fe$_{78}$Ni$_{22}$ thin films grown on Cu(100) substrates results in the localized transformation of the originally paramagnetic, face-centered-cubic continuous film into ferromagnetic patterns with body-centered-cubic structure. The direction of the magnetic easy axis can be controlled by the focused ion beam scanning strategy, resulting in eight differently oriented crystallographic domains with different magnetic properties. We study the local crystallographic orientations of the transformed areas by electron backscatter diffraction and correlate these results with local magnetometry measurements. The observed magnetic anisotropy can be explained as a result of residual lattice strain after the fcc$\to$bcc transformation. These results extend the understanding of this material system and its transformation and allow for the patterning of high-quality magnetic nanostructures with precisely controlled magnetization landscapes.}

\maketitle
\section*{Introduction}
Focused ion beam (FIB) irradiation \cite{Fassbender2008} or broad ion beam exposure through a lithographically defined mask \cite{Chappert1998, Rupp2008, Georgieva2007} provides a powerful and versatile approach for directly writing magnetic patterns, offering a rapid and flexible alternative to conventional lithography techniques for fabricating nanostructured magnetic samples~\cite{Bali2014, Roder2015, Mantion2022, Sorokin2023, Bunyaev2021, Urdiroz2021, Ehrler2020, Smekhova2024}. Metastable face-centered cubic (fcc) thin films of Fe or Fe-rich alloys \cite{Rupp2008, Zaman2011, Gloss2013} are good candidates for FIB magnetic patterning because they can be transformed by ion-beam irradiation to the ferromagnetic body-centered cubic (bcc) phase and the transformed patterns exhibit favorable magnetic properties, including high saturation magnetization, low damping, and the possibility to imprint magnetic anisotropy, making them ideal candidates for spin wave conduits \cite{Urbánek2018}. Recently, it has been demonstrated that it is possible to write magnonic waveguides by this technique, with fast spin-wave propagation at zero magnetic field by exploiting the fact that the imprinted magnetocrystalline anisotropy was strong enough to overcome the shape anisotropy of the waveguide \cite{Flajšman2020}. It is also possible to stabilize magnetic textures (magnetic vortices and domain walls) at desired positions and create magnonic devices, such as phase-shifters \cite{Wojewoda2020}.

\begin{figure}[b]
\includegraphics[width=0.450\textwidth]{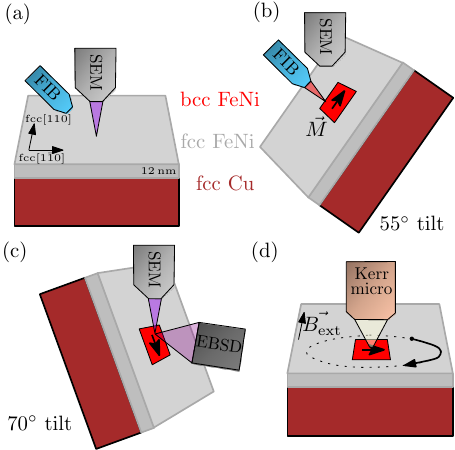}
\caption{\label{fig1}\textbf{A schematics of the transformation and characterization of the metastable FeNi film.} (a) Aligning the sample in SEM. (b) FIB transformation. (c) EBSD crystallographic analysis. (d) Kerr microscopy magnetic analysis.}
\end{figure}

However, further progress in complex device design requires a deeper understanding of the system, particularly in comprehending how the crystallographic properties of the transformed areas depend on the irradiation strategy and how they influence the magnetic behavior. The existing studies on the ion-irradiation induced fcc$\to$bcc transformation of Fe and Fe$_{x}$Ni$_{100-x}$ alloys primarily focus on analyzing the structural properties of films transformed by a broad argon ion beam with a relatively low energy of 5\,keV or less, using low-energy electron diffraction (LEED) \cite{Gloss2013} and scanning tunneling microscopy (STM) \cite{Rupp2008, Gloss2013}. The phase transformation process using a focused ion beam proceeds differently than the broad, low-energy ion beam transformation, as we have demonstrated in our previous work \cite{Urbánek2018}. A single-pass FIB scanning strategy provides local control over the recrystallization of individual areas with the possibility to control the magnetocrystalline anisotropy direction. However, the precise crystallographic configuration and the origin of the locally imprinted magnetic properties were not understood.

In this study, we explore the relationship between crystallographic structure and magnetic properties of FIB-written patterns. The patterns were written in 12\,nm-thick Fe$_{78}$Ni$_{22}$ films on Cu(001) using a 30\,keV Ga$^+$ FIB with a single-pass scanning strategy, with the scanning direction precisely aligned along the fcc[100] and fcc[010] crystallographic axes. These conditions ensured the highest crystallographic homogeneity in the transformed areas. After transformation, the patterns were structurally analyzed using high-resolution electron backscatter diffraction (EBSD) and magnetically characterized via magneto-optical Kerr effect magnetometry. The structural data were then used to model the magnetic behavior of individual crystallographic domains. Fig.~\ref{fig1} provides an overview of the experiments.  

\section*{Results}
\subsection*{EBSD crystallographic characterization}

\begin{figure}[t]
\includegraphics[width=0.450\textwidth]{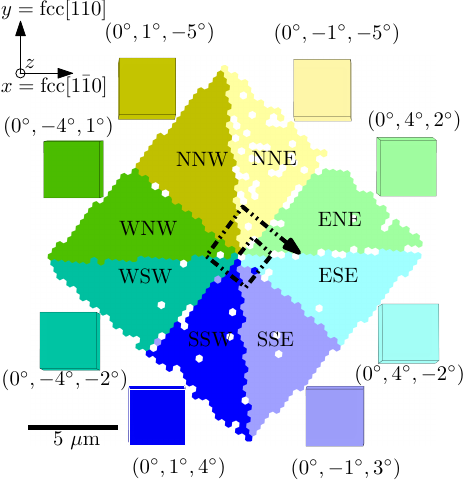}
\caption{\label{fig2} \textbf{Crystallographic EBSD measurement.} Magnetic 15×15\,µm$^2$ structure fabricated by 30\,keV single-scan FIB irradiation, which created eight bcc domains. These domains are named after the cardinal directions – for example NNE for north-northeast. The dashed-dot-dotted line represents the scanning strategy, which is inside-out and parallel to the fcc(100) directions. In the color-coded EBSD 2D map based on the IPF data, each color represents a specific crystallographic orientation. These orientations are depicted as oriented cubes in the vicinity of the corresponding domains. The rotation angles around the \mbox{($z$, $y$, $x$)} axes are in brackets.}
\end{figure}

An EBSD measurement of the transformed (bcc) area is shown in Fig.~\ref{fig2}. We processed the raw inverse pole figure (IPF) data into a color-coded EBSD 2D map  \cite{Randle1996}, where each color represents a specific crystallographic orientation. Eight bcc domains are clearly distinguishable. Four domain boundaries are located at the diagonals of the scanned square and run along the fcc[110] and fcc$[1\bar10]$ directions, where the focused ion beam took a 90$^\circ$ turn. Surprisingly, the other four domain boundaries appear at the midpoints of the square sides (running in the fcc[100] and fcc[010] directions). Here the FIB was scanning in a straight line; nevertheless the transformation resulted in two different structural domains. The crystal orientations are graphically depicted next to each domain as oriented unit-cell cubes together with the corresponding tilt angles. The tilt angles correspond to the chained rotation angles of the cube around \mbox{($z$, $y$, $x$)} axes. The maximum deviation from the reference orientation within each domain is  ±\,1.5$^\circ$, which indicates very good crystallographic quality and homogeneity of the transformed areas. We can see that each domain has a specific crystallographic orientation defined by one main tilt angle around the $x$ or $y$ axis (approx.\,4$^\circ$) and one minor tilt (1--2$^\circ$) around the other in-plane axis. The eight different crystallographic orientations are then given by the combinations of these two distinct tilts. Note that the opposing domains are tilted in opposite directions.

\begin{figure}[b!]
\includegraphics[width=0.450\textwidth]{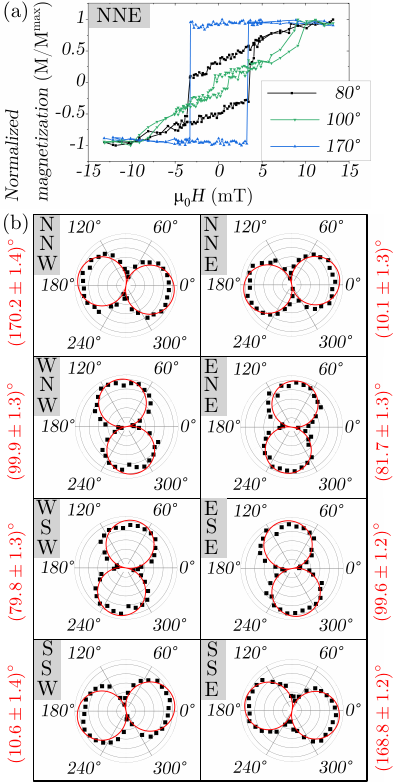}
\caption{\label{fig4}\textbf{Kerr microscopy measurements.} (a)~Typical hysteresis loops for three angles between external magnetic field and fcc$[1\bar{1}0]$ direction from the NNE domain. (b) Polar plots of the normalized remanence ${M_\mathrm{r}}/{M_\mathrm{r}^\mathrm{max}}$ for the eight domains (black rectangles). $0^\circ$ corresponds to the external field in the fcc$[1\bar{1}0]$ direction. Red lines represent fits according to the equation \ref{anisofit}. Fitted easy axis directions $\alpha_0$ are displayed by red numbers.}
\end{figure}

\subsection*{Local magnetometry characterization}
After completing the EBSD experiments, we transferred the sample to a magneto-optical Kerr microscope to perform local magnetometry measurements under ambient atmosphere. The domains were clearly visible in Kerr microscopy images due to~a different magnetization orientation in zero external magnetic field (see Supplementary Fig.~3). We acquired local magnetic hysteresis loops from each of the eight crystallographic domains, in an external magnetic field range of ±\,15\,mT. To evaluate the magnetic anisotropy, we first aligned the \mbox{fcc[110]} crystallographic direction with the direction of the external magnetic field. The plane of incidence of the reflected light was aligned parallel to the external magnetic field, thus the light polarization was sensitive to the projection of the magnetization into this direction (longitudinal Kerr effect). Then, we rotated the sample in 10$^\circ$ increments and acquired hysteresis loops. To obtain angle-dependent hysteresis loops, we examined the intensity variation of each of the eight domains (represented by grey levels proportional to magnetization) as a function of the external magnetic field. Fig.~\ref{fig4}a shows a typical example of three hysteresis loops that were acquired from the NNE domain at the three different angles of in-plane external magnetic field. 
From the hysteresis loops measured along the hard axis for all domains, an anisotropy field of $\mu_0H_\mathrm{ani}\approx9\,$mT was extracted using piecewise fitting. This corresponds to a uniaxial magnetic anisotropy constant $K_\mathrm{u}=\frac{\mu_0H_{\mathrm{ani}}M_\mathrm{S}}{2} = 6.3\cdot 10^{3}\,\mathrm{J/m^{3}}$ (assuming a saturation magnetization of $M_\mathrm{S}=1.4\cdot10^6\,\mathrm{A/m}$ \cite{Flajšman2020}).  

The easy axis orientations can be obtained from the angle-dependent remanent magnetization $M_\mathrm{r}(\alpha)$, where $\alpha$ is the angle between the external magnetic field and fcc$[1\bar{1}0]$. The results are presented in Fig.~\ref{fig4}b as individual polar plots for each structural domain, where black squares represent ${M_\mathrm{r}}/{M_\mathrm{S}}$. Then, we fitted the uniaxial anisotropy within each domain by the following formula \cite{Turcan2021, Kuschel2011}: 
\begin{equation}
\frac{M_\mathrm{r}}{M_\mathrm{S}}=\frac{M^{\mathrm{max}}_\mathrm{r}}{M_\mathrm{S}}\cdot \lvert\cos(\alpha-\alpha_0)\rvert +\frac{M^{\mathrm{off}}_\mathrm{r}}{M_\mathrm{S}}, 
\label{anisofit} 
\end{equation}
where $M^{\mathrm{max}}_\mathrm{r}$ is the maximum measured remanence, $M^{\mathrm{off}}_\mathrm{r}$ is a constant offset fixed to 0, and $\alpha_0$ is the direction of magnetic easy axis. 
The azimuthal easy axis orientation $\alpha_0$ (including uncertainty) for each domain is indicated by the red number adjacent to its polar plot.


\section*{Discussion}


\begin{figure}[t]
\includegraphics[width=0.48\textwidth]{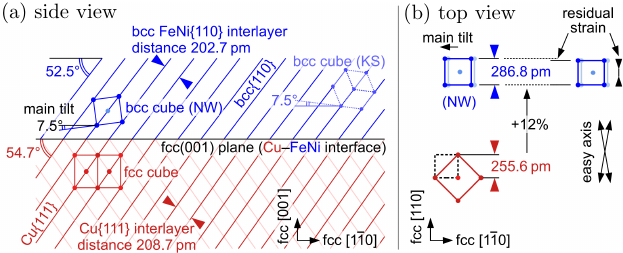}
\caption{\label{fig3} \textbf{Crystallography of bcc FeNi on fcc Cu.} The side view (a) shows the Nishiyama-Wassermann (NW) and Kurdjumov-Sachs (KS) orientational relationships, under the assumption that the fcc Cu\{111\} planes continue as bcc\{110\}. For the NW case, this would result in a 7.5° tilt of the bcc unit cell. The top view (b) shows that this transformation would require a large 12\% in-plane expansion along the fcc[110] direction. The direction of the main tilt corresponds to the WSW and WNW domains in Fig. \ref{fig2}.}
\end{figure}

The crystallographic structure of the transformed pattern, as observed by EBSD, can be explained by the Nishiyama-Wasserman orientation relationship \cite{nishiyama1934, wassermann1933}. The Nishiyama-Wasserman  (NW) relationship has parallel fcc\{111\} and bcc\{110\} planes (see Fig.~\ref{fig3}). In addition, NW is characterized by parallel fcc$\langle$110$\rangle$ and bcc$\langle$100$\rangle$  directions in the common plane. This orientation relationship of the in-plane directions differs from the Kurdjumov-Sachs (KS) orientation, which has the same parallel planes, but parallel fcc$\langle$110$\rangle$ and bcc$\langle$111$\rangle$ in-plane directions \cite{kurdjumow1930}. While NW results in a bcc film surface close to (001), KS orientation leads to a (tilted) bcc(110) surface. A KS-like orientation is observed after the fcc$\to$bcc transformation with low-energy Ar$^+$ ions \cite{Gloss2013} and also after the spontaneous fcc$\to$bcc transformation of ultrathin Fe/Cu(001) films \cite{Biedermann2001}.

In the current case of the transformation by 30\,keV Ga$^+$ irradiation, the near-(001) orientation of the bcc surface is a clear sign of the NW-orientation relationship. When assuming that the fcc Cu\{111\} atomic planes continue as bcc\{110\} as shown in Fig.~\ref{fig3}a, these bcc\{110\} planes would slightly deviate from the Cu\{111\} orientation (inclination 52.5 vs. 54.7°). Fig.~\ref{fig3}a also shows that the bottom of the bcc unit cell (the “bcc cube”) would be tilted by 7.5°. The tilt must lead to dislocations at the bcc-fcc interface. Thus, we attribute the lower tilt angle of  3--5° observed experimentally to the tendency to reduce the dislocation density. 

As seen in the top view in Fig.~\ref{fig3}b, the transformation requires a large 12\% lattice expansion along the fcc[110] direction, while the main tilt ensures that the strain in the perpendicular fcc[$1\bar{1}0$] direction is negligible. In contrast to the easy shear deformation in the fcc\{111\} or bcc\{110\} planes, this lattice expansion requires a high activation energy since it requires substantial mass transport out of the film (to the surface). Therefore, we consider it likely that full expansion will not occur, and the film will be under compressive stress in this direction. This stress may also provide a hint for explaining the unexpected domain boundaries in regions where the FIB writing direction does not change (e.g.,~between WNW and NNW in Fig.~\ref{fig2}). Switching between two almost perpendicular stress directions (two azimuthal directions of the tilt) reduces the associated elastic energy. At the turning points of the ion beam, the material transformed to bcc is surrounded by the (unstressed) fcc area in three (out of four) cardinal directions. Thus, the elastic energy will be lower, even without a large change in the tilt azimuth.

These crystallographic considerations explain the presence of four different azimuthal orientations of the main tilt direction (there are four different fcc\{111\} planes that may be converted to bcc\{110\}). Concerning the overall relation between the tilt directions and the FIB scanning scheme, we note that the main tilt direction is outward from the center where the ion irradiation starts and the eight domains nucleate (Fig.~\ref{fig2}). Thus, it is likely that the main tilt directions reflect the stress induced by ion irradiation, either directly (interstitials or other ion-beam-induced defects) or by the rapid recrystallization of the molten volume in the collision cascade \cite{Zaman2010, Busse2000}. This does not explain the slight~$\approx$1° minor tilt seen in the EBSD measurements. The minor tilt is likely related to the kinetics of the ion-induced transformation process.

The crystallographic picture outlined above predicts a substantial compressive strain along the main tilt axis (i.e. fcc[110] or fcc[1$\bar{1}$0]). This strain is expected to influence the magnetic properties. While shape anisotropy restricts the magnetization to in-plane directions, the origin of the uniaxial magnetic anisotropy observed by Kerr magnetometry can be interpreted as predominantly magnetoelastic. For in-plane magnetization, the magnetoelastic energy density term reads \cite{chikazumi1997physics}:
\begin{equation}
\begin{split}
E_{\mathrm{me}}(\alpha)
&=\frac{B_1}{2}\,(\varepsilon_{xx}-\varepsilon_{yy})\cos(2\alpha) \\
&\quad + B_2\,\varepsilon_{xy}\sin(2\alpha) + \mathrm{const.}
\end{split}
\end{equation}
Here $\alpha$ is the in-plane magnetization angle, $\varepsilon_{xx}$ and $\varepsilon_{yy}$ are the in-plane normal strain components, $\varepsilon_{xy}$ is the in-plane shear strain, and $B_1,B_2$ are magnetoelastic constants. 

The magnetic uniaxial anisotropy energy term $E_\mathrm{u}(\alpha)=K_\mathrm{u}\sin^2(\alpha-\alpha_0)$ (where $K_\mathrm{u}$ and $\alpha_0$ are obtained from the Kerr magnetometry measurement) can be matched to the magnetoelastic energy term $E_{\mathrm{me}}(\alpha)$ by equating the $\cos(2\alpha)$ and $\sin(2\alpha)$ coefficients, yielding the values of the normal strain difference $\Delta\varepsilon\equiv \varepsilon_{xx}-\varepsilon_{yy}$ and shear strain $\varepsilon_{xy}$:
\begin{equation}
\;\Delta\varepsilon = -\frac{K_u}{B_1}\cos(2\alpha_0)\;,
\;\varepsilon_{xy} = -\frac{K_u}{2B_2}\sin(2\alpha_0)\;.
\label{eq3}
\end{equation}
These determine the deviatoric in-plane strain tensor:
\begin{equation}
\varepsilon^{(\mathrm{dev})}=
\begin{pmatrix}
+\Delta\varepsilon/2 & \varepsilon_{xy}\\
\varepsilon_{xy} & -\Delta\varepsilon/2
\end{pmatrix}.
\end{equation}

The normal strain difference $\Delta\varepsilon$ is positive if the lattice is expanded in the $x$ direction and/or compressed in the $y$ direction. The symmetric shear component $\varepsilon_{xy}$ quantifies the change of the right angle between material lines initially along $x$ and $y$. 

From the crystallographic picture discussed above, it is clear that the normal strain difference $\Delta\varepsilon$ must be positive when the main tilt axis is $y$ (the lattice is compressed in the $y$ direction; this is the case of the WNW or WSW domain sketched in Fig.~\ref{fig3}). Vice versa, $\Delta\varepsilon$ must be negative when the main tilt axis is $x$. 

As the experiment shows, the easy axis is roughly parallel to the direction of the compressive strain, which implies that the magnetoelastic constant $B_1$ must be positive. Assuming a yield stress on the order of 150\,MPa and using the elastic stiffness coefficients from Ref.~\cite{Leese1968}, we estimate a normal-strain difference of $\Delta\varepsilon \approx 10^{-3}$. Combining this value with the Kerr-derived anisotropy then yields $B_1 \approx +6\cdot10^{6}\,\mathrm{J\,m^{-3}}$, which is a typical magnitude for magnetoelastic constants.

Note that in pure bcc Fe the magnetoelastic constant $B_1$ is negative~\cite{burkert_calculation_2004}. However, the magnetoelastic constants depend sensitively on the electronic structure. Alloying with a few percent of Co, which increases the d-band filling, leads to a sign reversal of the magnetostriction~\cite{jen1999magnetostriction}. Since Ni also has more d electrons than Fe, a similar sign reversal can be expected for our Fe$_{78}$Ni$_{22}$ alloy.

Since the magnetoelastic constants $B_1$ and $B_2$ of the Fe$_{78}$Ni$_{22}$ alloy are not known, we report the magnetoelastic strain products $B_1\Delta\varepsilon$ and $B_2\varepsilon_{xy}$, calculated from the Kerr-derived $\alpha_0$ and $K_\mathrm{u}$ using Eq.~(\ref{eq3}). The values for each structural domain (defined in Fig.~\ref{fig2}) are summarized in Table~\ref{table1} together with the EBSD tilt angles and the Kerr easy-axis orientations.
\begin{figure}[b]
  \centering
    \centering
    \setlength{\tabcolsep}{3pt}
    \small
    \begin{tabular}{lcccccc}
        \hline
        domain & \makecell{Kerr \\ $\alpha_0[^\circ]$} & \makecell{$\omega_x$\\$[^\circ]$} & \makecell{$\omega_y$\\$[^\circ]$} & \makecell{EBSD\\tilt axis} & \makecell{$B_1\Delta\varepsilon$\\$[\mathrm{kJ/m^{3}}]$} & \makecell{$B_2\varepsilon_{xy}$\\$[\mathrm{kJ/m^{3}}]$}\\[1em]
        \hline
        NNW & 170 & -5 & +1 & x & -5.9 & +1.1 \\
        NNE & 10 & -5 & -1 & x & -5.9 & -1.1 \\
        ENE & 82 & +2 & +4 & y & +6.0 & -0.9 \\
        ESE & 100 & -2 & +4 & y & +5.9 & +1.0 \\
        SSE & 169 & +3 & -1 & x & -5.8 & +1.2 \\
        SSW & 11 & +4 & +1 & x & -5.9 & -1.1 \\
        WSW & 80 & -2 & -4 & y & +5.9 & -1.1 \\
        WNW & 100 & +1 & -4 & y & +5.9 & +1.1 \\
        \hline
    \end{tabular}  
    \captionof{table}{\label{table1} \textbf{Summary of magnetic and crystallographic parameters.} For each domain, the table displays the fitted magnetic easy axis (Kerr EA), rotation angles around the $x$ and $y$ axes ($\omega_x, \omega_y$), the designated major rotation axis, and calculated values of $B_1\Delta\varepsilon$ and $B_2\varepsilon_{xy}$. Strains can be obtained by dividing by material-specific constants $B_1$, $B_2$.}
\end{figure}
Table~\ref{table1} reveals two robust correlations between crystallography and magnetism. First, the normal-strain difference $\Delta\varepsilon=\varepsilon_{xx}-\varepsilon_{yy}$ changes sign when the main EBSD tilt axis switches between $x$ and $y$, which is consistent with the picture of the strain along the main tilt axis being the dominant contribution to the magnetic anisotropy. Second, the experimentally observed $\pm10^\circ$ deviations of the magnetic easy axis from the cardinal directions cannot be reproduced by $\Delta\varepsilon$ alone and require a shear component $\varepsilon_{xy}$. The shear contribution reverses between domains written with perpendicular FIB scanning directions, which is reflected by the sign changes of $B_2\varepsilon_{xy}$ in Table~\ref{table1}.

To obtain an order-of-magnitude estimate of the actual shear distortion, one may assume $B_2=1\cdot10^{6}\,\mathrm{J/m^{3}}$, which yields $\varepsilon_{xy}\approx 10^{-3}$, corresponding to a shear angle of approximately $\gamma\approx 0.1^\circ$. For larger values of $B_2$, which may reach a few times $10^{7}\,\mathrm{J/m^{3}}$, the corresponding shear distortion would be proportionally smaller. In any case, the expected shear distortion is below the angular resolution of EBSD, analogous to the compressive strain component.

Since the sign of $B_2$ for Fe$_{78}$Ni$_{22}$ is not known, the sign of the shear strain cannot be determined. From the Kerr measurements, however, it follows that the shear component must change sign between domains with perpendicular FIB scanning directions (e.g.\ WNW and WSW), which is reflected in the sign changes of $B_2\varepsilon_{xy}$ summarized in Table~\ref{table1}. This correlation suggests that the shear strain and the small $\approx 1^\circ$ minor tilt observed by EBSD (which is also opposite for WNW and WSW) have a common origin.



\section*{Conclusion}
In summary, we have presented a metastable Fe-Ni alloy suitable for rapid prototyping of magnetic structures via FIB direct writing. We have shown that eight different, highly homogeneous structural domains can be created by scanning with the FIB along specific fcc$\langle100\rangle$ directions. We performed crystallographic analysis by EBSD and described the influence of the FIB writing scheme on the crystallographic domains. Our data indicate that the observed in-plane uniaxial magnetic anisotropy originates from magnetoelastic coupling associated with residual strain generated during the FIB-induced fcc$\to$bcc phase transformation. By combining Kerr magnetometry with magnetoelastic modeling, we 
establish a clear correlation between the FIB scanning strategy and the resulting anisotropic strain components. This strain-mediated control provides a mechanism for tailoring the magnetic easy axis through appropriate selection of the FIB writing conditions. This understanding of the connection between structural and magnetic properties of metastable Fe-Ni alloy films undergoing ion-induced transformation opens new possibilities to design materials with properties unachievable in magnetic thin films patterned by other methods. Given the low damping and good spin wave propagation characteristics \cite{Flajšman2020}, such a system can be exploited, for example, in magnonics applications \cite{Wang2024}.  

\backmatter

\section*{Methods}
\textbf{Metastable FeNi film growth.}
The studied films were grown in an ultrahigh vacuum (UHV) system via electron-beam evaporation from a Fe$_{78}$Ni$_{22}$ alloy source (2\,mm thick rod, 99.99\,$\%$ purity), following the procedure described in \cite{Gloss2013}. Prior to deposition, the Cu(001) substrates were prepared by multiple cycles of sputtering with 2\,keV Ar$^+$ ions for 30 minutes and annealing at 600\,$^\circ$C for 10 minutes to ensure surface cleanliness. The cleanliness of the surface and the film composition were verified using Auger electron spectroscopy (AES).

The deposition was carried out at a pressure of $5 \cdot 10^{-10}$\,mbar with a calibrated deposition rate of 0.02\,\AA/s, as determined by a quartz-crystal microbalance. For the 12\,nm films used in this study, the deposition time was approximately 90 minutes. To minimize the influence of high-energy ions from the evaporator \cite{Nagl1994}, which could lead to an unwanted fcc $\to$ bcc transformation, a repelling voltage of +1.5\,kV was applied to a cylindrical electrode positioned at the evaporator orifice.

\textbf{FIB patterning.}
The FIB-induced magnetic patterning was conducted using a combined focused ion beam–scanning electron microscope system (FIB-SEM, Lyra3, Tescan) after transferring the sample through ambient atmosphere. The experimental sequence began by aligning the sample’s \mbox{fcc$[1\bar{1}0]$} and \mbox{fcc[110]} crystallographic axes to the SEM x-y coordinate system to define the FIB scanning direction. This alignment was achieved using a pseudo-Kikuchi diffraction pattern \cite{Baba2002} visible in low-magnification SEM images (see Supplementary Fig.~1a).

After alignment, the sample stage was tilted by 55$^\circ$ towards the FIB column to ensure a perpendicular angle of incidence for the 30\,keV Ga$^+$ ions. The FIB current was set to 160\,pA, with a spot size of 30\,nm and a step size of 10\,nm. The resulting ion dose was $10^{15}\,\mathrm{ions}/\mathrm{cm}^2$. During the transformation, the focused ion beam scanned the sample surface in a square spiral pattern, starting from the center of a 15×15\,µm$^2$ square (dashed-dot-dotted line in Fig.~\ref{fig2}).

The transformed region was immediately visible in secondary electron images, revealing crystallographic contrast consisting of eight structural domains. This contrast, which most likely arises from the channeling of primary electrons, was most pronounced at a stage tilt angle of $\pm5^\circ$ (see Supplementary Fig.~1b).

\textbf{EBSD measurement.}
After patterning, the samples were transferred to a high-resolution scanning electron microscope (SEM) equipped with an electron backscatter diffraction (EBSD) detector (FEI Verios 460L with an EDAX DigiView IV insertable camera). The EBSD measurements were conducted to determine the crystallographic orientation of individual domains within the transformed region. The measurements were performed with an electron beam voltage of 15\,keV and a beam current of 1.6\,nA, while maintaining a working distance of approximately 9\,mm. The sample was tilted by 70° from the horizontal (corresponding to 20° relative to the incident electron beam), while the EBSD detector was positioned below the sample to collect the diffracted electrons. Data acquisition was carried out over the 15 × 15\,µm$^2$ transformed area, with a total measurement time of approximately 2.5 hours.

To enhance the visibility of Kikuchi patterns from the 12\,nm-thick layer, despite the strong diffraction signal from the underlying substrate, dynamic background division followed by intensity histogram normalization was applied. Post-processing of the EBSD data was performed using OIM Analysis™ software \cite{OIM}, incorporating a confidence index (CI)-based clean-up procedure. This included non-destructive Grain CI standardization, Neighbor Orientation Correlation, and Grain Dilation with a CI threshold value of 0.1. The processed Kikuchi patterns, 2D phase maps, and inverse pole figure (IPF) orientation map are shown in Supplementary Fig.~2.

\textbf{Local magnetometry.}
We used a magneto-optical Kerr microscope (Evico Magnetics) for local magnetometry. Magnetic hysteresis loops were acquired from each of the eight crystallographic domains within a magnetic field range of ±15 mT. The sample was rotated in 10$^\circ$ increments to obtain angle-dependent hysteresis loops. The intensity variation of each domain was analyzed as a function of the magnetic field using a LabVIEW program \cite{githubKerr} (see Supplementary Fig.~3).

\bmhead{Acknowledgments}
This research was supported by the project No. CZ.02.01.01\slash00\slash22\textunderscore008\slash0004594 (TERAFIT) and by the joint project of Grant Agency of the Czech
Republic (Project No. 15-34632L) and Austrian Science Fund (Project No. I 1937-N20). CzechNanoLab project LM2023051 is acknowledged for the financial support of the measurements and sample fabrication at CEITEC Nano Research Infrastructure.








\bmhead{Authors' contributions}
J.H. grew metastable films under supervision of M.S, J.H. performed the FIB transformation. J.H. and O.W. ran Kerr and EBSD characterization. J.H. and M.S. evaluated EBSD data, J.H., O.W. and L.V. evaluated magnetic data. L.V. and M.U. performed magnetic modeling. J.H. wrote the manuscript with supervision of M.U. All contributed in writing and reviewing the paper.

\bmhead{Competing interests}

The authors declare no competing interests.

\bmhead{Data and code availability}
All data and code used to generate the presented figures are
available in the Zenodo repository (DOI:10.5281/zenodo.15729695).



%
%
%
%

\bibliography{sn-bibliography}
\begin{appendices}





\end{appendices}




\end{document}


\title[Article Title]{Supplementary information: Precise control of crystallography and magnetism in focused-ion-beam transformed iron-nickel thin films}


\author*[1]{Jakub Holobr\'{a}dek} \email{holobradek@vutbr.cz}

\author[1]{Libor Voj\'{a}\v{c}ek}

\author[1]{Ond\v{r}ej Wojewoda}

\author[3]{Michael Schmid}

\author*[1,2]{Michal Urb\'{a}nek}  \email{michal.urbanek@ceitec.vutbr.cz}

\affil[1]{\orgdiv{CEITEC BUT}, \orgname{Brno University of Technology}, \orgaddress{ \city{Brno}, \country{Czech Republic}}}

\affil[2]{\orgdiv{Institute of Physical Engineering}, \orgname{Brno University of Technology}, \orgaddress{\city{Brno}, \country{Czech Republic}}}

\affil[3]{\orgdiv{Institute of Applied Physics}, \orgname{TU Wien},  \city{Vienna}, \country{Austria}}

\maketitle

\renewcommand\figurename{Supplementary Fig.}

\begin{figure*}[h]
\centering
\label{S1}
\includegraphics{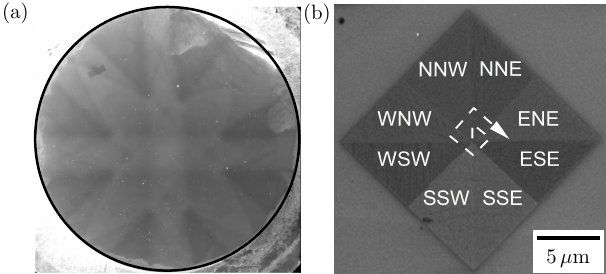}
\caption{(a) Pseudo-Kikuchi diffraction pattern used for sample alignment in SEM. (b) SEM micrograph of a magnetic 15×15\,µm$^2$ structure fabricated by 30\,keV single-scan FIB irradiation with 5$^\circ$ microscope stage tilt. The dashed line represents the scanning strategy. The scanning strategy is inside-out and parallel to the fcc(100) directions. In the image contrast, eight domains are visible. These domains are named after the cardinal directions – for example NNE for north-northeast.}
\end{figure*}

\begin{figure*}[h]
\centering
\label{S2}
\includegraphics{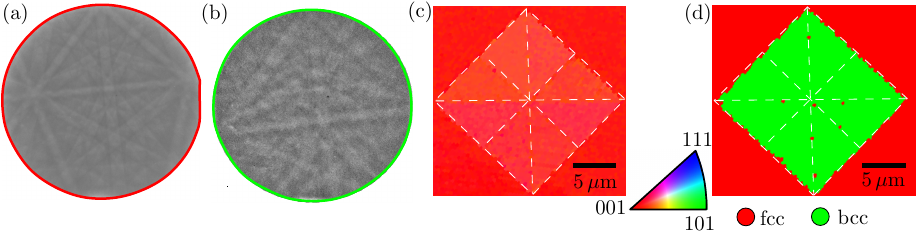}
\caption{(a) EBSD Kikuchi patterns after background subtraction of the as-grown fcc FeNi film and (b) the FeNi film after transformation to bcc. (c) Typical EBSD Inverse pole figure (IPF) orientation map of the bcc structure from Fig. S1b after CI cleanup. The irradiated areas are very close to (001) orientation. White dashed lines indicate eight domains formed after the transformation. (d) Phase fcc/bcc map of the structure.}
\end{figure*}


\begin{figure*}[h]
\centering
\label{S3}
\includegraphics{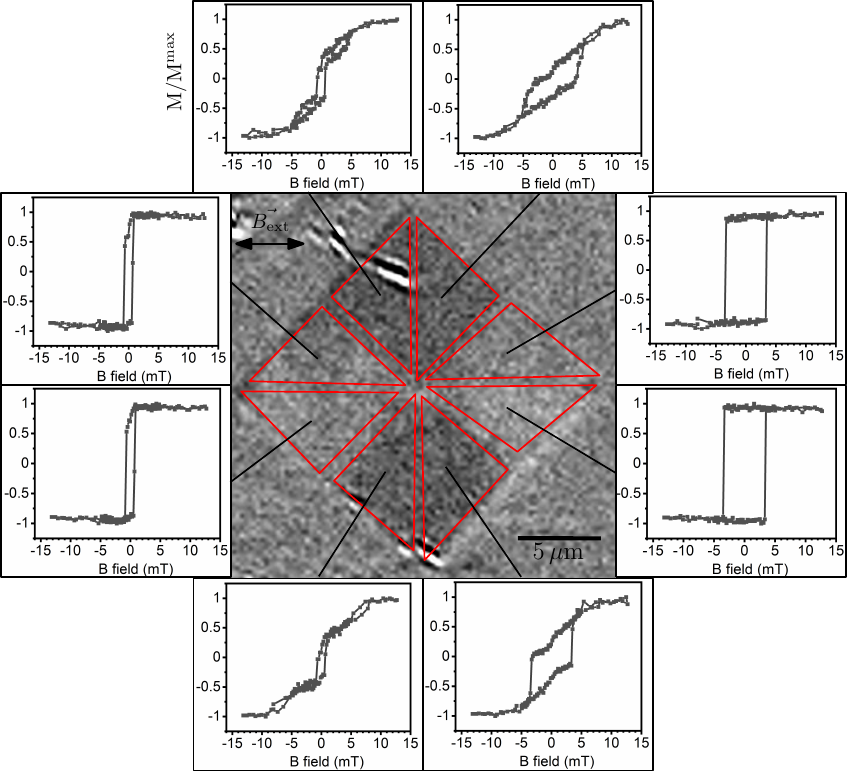}
\caption{Postprocessing of Kerr image in the LabVIEW program. Hysteresis loops were extracted from the red regions. Example of the Kerr image taken in zero external field.}
\end{figure*}



\clearpage